\documentclass[aps,pra,twocolumn,showpacs,superscriptaddress,10pt,letterpaper,floatfix]{revtex4-1}
\usepackage{graphicx}% Include figure files
\usepackage{dcolumn}% Align table columns on decimal point
\usepackage{bm}% bold math
\usepackage{color}% add textcolor
\usepackage{amsmath}% for subequations environment
\usepackage{mdframed}
\usepackage{enumitem}
\usepackage[caption=false]{subfig}
\usepackage{mathtools}
\usepackage{tabularx}
\usepackage{multirow}
\usepackage{appendix}
\usepackage{longtable}

\definecolor{red}{rgb}{1,0,0}
\definecolor{orange}{rgb}{1,0.5,0}
\definecolor{green}{rgb}{0.13,0.55,0.13}
\definecolor{purple}{rgb}{0.5,0,1}

\begin{document}

\title{Identifying features predictive of faculty integrating computation into physics courses}

\author{Nicholas T. \surname{Young}}
\affiliation {Department of Physics and Astronomy, Michigan State University, East Lansing, Michigan 48824}
\author{Grant \surname{Allen}}
\affiliation {Department of Physics and Astronomy, Michigan State University, East Lansing, Michigan 48824}
\author{John M. \surname{Aiken}}
\affiliation {Department of Physics and Astronomy, Michigan State University, East Lansing, Michigan 48824}
\affiliation {Center for Computing in Science Education \& Department of Physics, University of Oslo, N-0316 Oslo, Norway}
\author{Rachel \surname{Henderson}}
\affiliation {Department of Physics and Astronomy, Michigan State University, East Lansing, Michigan 48824}
\author{Marcos D. \surname{Caballero}}
\email[Corresponding Author: ]{caballero@pa.msu.edu}
\affiliation {Department of Physics and Astronomy, Michigan State University, East Lansing, Michigan 48824}
\affiliation {Center for Computing in Science Education \& Department of Physics, University of Oslo, N-0316 Oslo, Norway}
\affiliation {CREATE for STEM Institute, Michigan State University, East Lansing, Michigan 48824}

\date{\today}

\begin{abstract}
Computation is a central aspect of 21st century physics practice; it is used to model complicated systems, to simulate impossible experiments, and to analyze mountains of data. Physics departments and their faculty are increasingly recognizing the importance of teaching computation to their students. We recently completed a national survey of faculty in physics departments to understand the state of computational instruction and the factors that underlie that instruction. The data collected from the faculty responding to the survey included a variety of scales, binary questions, and numerical responses. We then used Random Forest, a supervised learning technique, to explore the factors that are most predictive of whether a faculty member decides to include computation in their physics courses. We find that experience using computation with students in their research, or lack thereof and various personal beliefs to be most predictive of a faculty member having experience teaching computation. Interestingly, we find demographic and departmental factors to be less useful factors in our model. The results of this study inform future efforts to promote greater integration of computation into the physics curriculum as well as comment on the current state of computational instruction across the United States.
\end{abstract}

% PACS, the Physics and Astronomy Classification Scheme.
%\keywords{\textcolor{red}{Physics education research, mathematics, upper-division}}%Use showkeys class option if keyword display desired
\maketitle

\section{Introduction\label{sec:intro}}

Computation is a central practice of modern scientific research that has enabled numerous experimental and theoretical discoveries in physics. While this practice is part and parcel to the work of modern physicists, it is not often represented in physics curriculum \cite{Fuller:2006wg,Chonacky:2008gq,caballero_prevalence_2018}. This is despite the current push from various professional and governmental organizations for the integration of computation into a variety of fields and contexts, including physics \cite{NGSS, PCAST, behringer_guest_2017}. Integrating computation into physics courses represents a shift in the curriculum and, thus, requires faculty to develop, adopt, or adapt materials appropriate for their courses and students. To support faculty and further integration efforts, we need to understand why faculty choose to integrate computation into their courses or why they choose not to do so. In this paper, we address this issue by determining which factors are predictive of a physics faculty member having experience teaching computation to undergraduate students.

As computation is up and coming as an instructional tool and strategy in physics, there is little literature on the experiences that faculty have when integrating computation into their courses. While we expect there to be challenges unique to integrating computation into the physics classroom, we also expect faculty to encounter similar difficulties as they would for implementing other instructional efforts including research-based instructional strategies (RBIS)  \cite{henderson_barriers_2007,turpen_perceived_2016}. For example, faculty may be concerned about having to make time to teach students basic programming principles in addition to the physics content they are already required to cover or having to create instructional materials that utilize computation. Work on faculty change has found that these concerns may be alleviated by supporting and encouraging faculty as they implement RBIS \cite{henderson_use_2012,turpen_perceived_2016}. The Partnership for Integration of Computation into Undergraduate Physics (PICUP) is one such group currently working to support faculty as they integrate computation into their courses \cite{PICUP}. However, to alleviate such concerns, we first must understand the nature of those concerns and how they might impact whether a faculty member uses computation in their classroom and why some faculty have decided to include computation in their classrooms. In this paper, we use a machine learning technique called Random Forest to address the latter, that is to determine the factors that are predictive of whether a faculty member has experience teaching computation or not.

This paper is organized as follows: Sec.~\ref{sec:background} provides an overview of the findings from institutional change literature in physics. We then provide an overview of the Random Forest methodology and its implementation in Sec.~\ref{sec:methodology} with details appearing in a supplemental material. In Sec.~\ref{sec:results}, we describe our findings followed by the work done to validate those findings (Sec.~\ref{sec:results}) and the resulting limitations of those findings (Sec.~\ref{sec:limitations}). We conclude with a discussion our findings (Sec.~\ref{sec:discussion}) and their broader implications (Sec.~\ref{sec:conclusions}).

\section{Background}\label{sec:background}

To understand what those concerns or factors that might affect a faculty member's decision to teach computation to their students, we can look at which factors are predictive of a faculty member trying RBIS, continuing to use RBIS, and using multiple RBIS. Because we are interested in whether faculty have experience teaching computation, we are interested in whether faculty have reached the implementation stage of Rogers' five stages for adopting an innovation, which include knowing of the innovation, becoming persuaded to adopt the innovation, deciding to adopt the innovation, implementing the innovation, and continuing to use the innovation \cite{rogers1995diffusion}. We leave questions about why faculty continue or do not continue to use computation to future work. Prior work into adopting new instructional strategies suggests that faculty choose to implement RBIS based on their own decisions \cite{foertsch1997persuading, ho2001conceptual,ellsworth2000surviving,Henderson2005challenges, prosser1999understanding}. Henderson, Dancy, and Niewiadomska-Bugaj extended this line of work by looking at specific factors predictive of adopting a new instructional strategy and found that regularly reading teaching journals, attending talks and workshops related to teaching, attending the New Faculty Workshop, having an interest in using more RBIS, and the type of institution (two-year college, four-year bachelor granting institution, and four-year doctoral granting institution) are predictive of a faculty member trying an RBIS \cite{henderson_use_2012}. Alternatively, they found that factors such as class size, research productivity, job type (lecturer, full professor, etc.), departmental encouragement, years of teaching experience, course type (algebra-based or calculus-based), and demographic factors such as gender and highest degree obtained were not predictive of a faculty member trying a RBIS. In addition, perceived implementation challenges such as situational characteristics, including resistance from students, large of class sizes, pressure to cover a large amount of content in the course, and student expectations about how class should be structured, and personal reasons, such as the perceived amount of time it would take to implement the change and having a bad experience with trying to implement the change, could prevent a faculty member from trying to implement a RBIS \cite{henderson_barriers_2007,turpen_perceived_2016}.

Even though computation is a technique to do physics rather than an instructional strategy like RBIS, its use is informed by physics education research and it is not typically found in traditional-lecture based courses. Just as adopting a RBIS requires adopting new ways, tools, and methods of teaching, integrating computation into a course or curriculum also requires adopting these. Further, many of the implementation challenges for RBIS, such as not enough time to fit in new content, the amount of  time needed to implement the change, and student resistance, have been also documented for computation when a department tries to transform their courses to include computation \cite{taylor_using_2006, johnston_implementing_2006, martin_undergraduate_2017}. Therefore, we believe that we can treat computation like a research-based instructional strategy (RBIS) at least with respect to adoption and implementation. We can apply an institutional change lens to interpret our results and, thus, we expect that the features we find to be predictive of whether faculty have experience teaching computation to also be important predictors of whether faculty have implemented RBIS as found in the literature.

\section{Methodology\label{sec:methodology}}
The factors found in the literature and the data we use in this study consist of binary, Likert-scale, and open response questions. While these data are not uncommon in physics education research (PER), they are often themselves the only source of the data (i.e., only one form of response) or they are part of some larger data set where some number of the same type of response formats (e.g., multiple choice responses with a single correct answer) are the main source of data. In both cases, PER has accepted methods for analyzing this kind of data to determine key factors that predict the outcomes most strongly. For example, when these data are part of a larger data set, a regression analysis can be performed where the categorical data are treated using binary codes (i.e., ``dummy'' variables) and are then included in the regression analysis. This is common technique used in a number of studies in PER \cite{hazari2010connecting,henderson_use_2012,heron2015effect}. However, performing such a regression analysis on our data is problematic. As most of our data take some categorical form, our data violates key assumptions in any linear regression model such as normality and equal spreads. \cite{fox1997applied,draper2014applied}. As such, the questions posed in our work are rooted in a classification task: what features predict which faculty have experience teaching computation and those that do not? Characterizing our study as a classification task led us to employ a supervised learning method appropriate for the data -- the Random Forest algorithm.
Below, we provide a brief overview of the algorithm and our implementation specific to this study (Sec.~\ref{subsec:implementation}). In the supplemental material, we provide more thorough background on the Random Forest algorithm including how the model is validated (Sup.~Sec.~II), how a subset of important features can be identified (Sup.~Sec.~III), and how to handle bias in the model (Sup.~Sec.~IV).

\subsection{The Random Forest algorithm}\label{subsec:rf}

A Random Forest is a supervised machine learning approach that can be used to classify data into categories or model outcomes over some range using regression. The algorithm can also be used to develop a quantitative, relative measure of how important certain factors are in predicting those categories or outcomes. \cite{ho1995random,ho1998random,breiman2001random}. As with all supervised machine learning techniques, a Random Forest is trained on a data set with known classifications. The model is grown using binary decision trees and the results of each decision tree is aggregated into a single result \cite{rokach2014data, breiman2001random}. The randomness comes from the fact that only a fraction of the factors are used to construct the decision trees and only a fraction of the data, controlled by the training fraction is used to test the model (see supplemental material). Through this training, the algorithm develops a model for the data set. Then, the model is applied to a set of sequestered data, known as the test set, that was not used in the original training. The model is used to predict the classifications for this testing data, which are also known. 

In order to assess the model, a few measures are employed. First, the accuracy of the model is computed, which is the fraction of the data in the test set that was correctly classified. Second, a receiver operating characteristic curve (ROC curve) of the model is generated \cite{fawcett2006introduction}. An example is shown in Fig. 2 of the supplemental material. The ROC curve plots the true positive rate (proportion of people who have a specific trait such as teaching computation that are correctly classified as having that trait) as a function of the false positive rate (the proportion of people who do not have that specific trait but are incorrectly classified as having the trait). The ROC curve allows one to visualize the trade-off between creating a model that has a high true positive rate but has many false positives and models have few false positives but also fewer true positives. The ROC curve can be represented as a single number called the area under the curve (AUC), where a perfect classifier will have an AUC of 1 while a binary classifier that is randomly classifying data will have an AUC of 0.5. While there is not general agreement on what constitutes different levels of the significance for the AUC measure, the literature suggests that an AUC $>$ 0.7 is a reasonable lower bound for a random forest model \cite{araujo2005validation}.

In addition to classifying data, the Random Forest algorithm is able to empirically determine the relative importance of each factor to the model. While there are many ways to calculate this importance, we chose to use an importance measure based on the AUC because it  has been shown to be unbiased when the predicted variable is unbalanced and the data types of the factors are different \cite{janitza_auc-based_2013} as is the case with our data. Using this importance measure, the relative importance of a factor to the model is determined how much the AUC changes when the information from that factor is removed from the model. If the factor removed from the model was useful for prediction, the AUC will decrease to a greater extent than if the factor was less useful for making predictions. By removing each factor one at a time, the change in AUC can be determined for each factor and the relative importance of each factor can be determined.

However, computing the relative importance for each factor does not provide any information about whether the factor is actually important to the model. To determine this, some type of selection technique must be used. While many techniques exist, we used recursive backward elimination, which means the less important features were recursively removed from the model until the ``best'' model is found as measured by the accuracy \cite{diaz-uriarte_gene_2006}. Here, ``best'' model refers to the model that uses the fewest number of factors to produce a model that is within 1 standard error of the highest possible accuracy.

%\subsubsection{Other machine learning approaches}

Random forests are one of a number of different possible machine learning classifiers that can be used on any given data set. Na\"{i}ve-Bayes methods, support vector classifiers, k-nearest neighbors, and gradient tree boosting are all possible classification schemes that could have been used for this study. Olson and collaborators modeled a number of open-source data sets with these and other classifiers in order to offer best practices for using machine learning classification algorithms \cite{olson2017data}. In their work, Olson \textit{et al}. found the Random Forest algorithm to be one of the best algorithms overall -- second only to gradient tree boosting. In head-to-head comparisons where parameter tuning was allowed, Random Forest predictions were as accurate, within error, of gradient tree boosting and support vector machines. In principle, several algorithms could be applied to the same data set and the resulting classifications compared, but that is not the purpose of our work. We selected the Random Forest algorithm for its documented robustness and its intuitive nature.

\subsection{Implementation\label{subsec:implementation}}

\subsubsection{Survey}\label{subsubsec:Survey}

To determine which factors are most important in predicting whether a faculty member has experience teaching computation, we analyzed survey responses from 1246 faculty at 357 unique institutions \cite{caballero_prevalence_2018}. The survey focused around five broad topics: attitudes toward computation, experience with computation, computational resources provided by their department, motivations for teaching or not teaching computation, and departmental views of computation. As prior work on adopting research-based instructional strategies has found that learning about new strategies as well as interest in using new RBIS to be significant explanatory variables in determining whether a faculty member will try a new RBIS \cite{henderson_use_2012}, we expect that faculty's attitudes toward computation will be predictive of them choosing to incorporate it into their classroom. For example, we would expect a faculty member who sees clear benefits of using computation in the classroom such as allowing for new problems and concepts to be covered in the course or being able to visualize or simulate phenomena to incorporate computation into the classroom. Likewise, we would expect that a faculty member who has experience with computation, either having learned it during their schooling or using computation in research or other non-teaching duties, would be more likely to integrate computation into their classroom than an instructor who has never used computation before and hence would have to teach themselves before including computation in their classrooms. While Henderson, Dancy, and Niewiadomska-Bugaj did not find departmental encouragement or research productivity measures to be useful explanatory variables for determining whether a faculty member tried a RBIS \cite{henderson_use_2012}, faculty might be motivated to incorporate computation if their department encourages them with incentives for integrating computation into their courses (such as increased resources for the course or as criteria for tenure/promotion) or if they believe using computation in their courses would open new funding opportunities or other research benefits. Finally, demographic or institutional factors may influence a faculty member's willingness to incorporate computation into their course and therefore, questions regarding these factors were also included on the survey.

The constructed survey items varied in scale of measurement from yes/no-binary questions, to Likert scales and open-ended responses. Given the broad range of questions and the fact that not all questions would be relevant to all survey takers, the survey used binary logic; some survey participants saw different questions based on their response to the first question, ``do you have experience teaching computation". Of the 1246 respondents, 751 faculty said they did have experience teaching computation while 495 faculty said they did not have experience teaching computation. 

\subsubsection{Sample}\label{subsubsec:Sample}

In order to determine important factors for integrating computation into physics courses, we could only use questions that were seen by both faculty who have and do not have experience teaching computation. This left us with 44 questions which were binary, Likert-scale, and open response. Because the 44 questions we selected were of different data types (from binary to near continuous) and our data set was unbalanced, we utilized conditional inference forests via the \texttt{cforest} function in the Party package for R \cite{Rprogram, hothorn_unbiased_2006,strobl_bias_2007, strobl_conditional_2008}. As Kim and Loh have found that different proportions of missing values can introduce bias into classification trees, we excluded any faculty member from the sample who did not answer all 44 questions \cite{kim_classification_2001}. We address our choice to remove faculty who did not answer all questions from the sample instead of using multiple imputations or other methods to address missing data in Sec.~\ref{sec:limitations}. After doing this procedure, we were left with 693 faculty (56\% of our sample). In the original sample, 60\% of the faculty had indicated that they had experience teaching computation while in our reduced set with only faculty who answered all questions, 62\% indicated they had experience teaching computation, suggesting that the data we are using is still representative of the overall sample.

\subsubsection{Growing the Random Forest}\label{subsubsec:GRF}
To run the cforest algorithm, we first randomly split the data into a training set and a testing set, where 70\% of the data was used in the training set (corresponding to a training fraction of .70), a common value in the literature \cite{polat_detection_2007,chen_comparative_2017,ahneman_predicting_2018}. Next, we set mtry=$\sqrt{N}$ using \texttt{cforest\_control}, where ntree is the default value in the \texttt{cforest} algorithm and mtry is equivalent to $n_{in}$ in \cite{svetnik2003random}. All other cforest function parameters were set to their default values from \texttt{cforest\_unbiased}, which includes subsampling without replacement and $n_{tree}=500$. We then ran the cforest algorithm on our data set to grow the forest. To calculate the accuracy and AUC of the model, we used the \texttt{caret} and \texttt{ROCR} packages \cite{caret, ROCR}. To calculate the variable importances we used the \texttt{varimpAUC} function from the party package. We then ran the \texttt{cforest} algorithm an additional 29 times, for a total of 30 trials, allowing us to use the central limit theorem \cite{CLT} to define the mean and standard error of the importances. Thirty trials is typically the minimum number of trials to apply the central limit theorem \cite{hogg2015probability} and Shapiro-Wilk tests, a test of normality where the null hypothesis is that the data are normally distributed, show that with 30 trials, the data are normally distributed; therefore, additional trials were deemed not necessary and would have only consumed additional computational resources. Further, QQ plots \cite{wilk_probability_1968}, which are scatterplots that compare the theoretical normal distribution with the actual data and will appear  as lines if the data are in fact linear, do not show any nonlinear behavior, suggesting that 30 trials was sufficient. In addition, the data were re-split into training and testing data sets before each trial to minimize the inherent randomness of the \texttt{cforest} algorithm and any bias that could result from the training data not being representative of the overall data.

In order to find the meaningful factors, we used the recursive backward elimination technique described in Ref.~\cite{diaz-uriarte_gene_2006} as we did not want to presuppose a set number of meaningful variables, only had a few negative values such that the resulting distribution would not be useful, and generating null importances for our 44 variables would not be practical (see supplemental material for details of these alternative approaches). The recursive backward elimination technique was implemented through a modification of \texttt{varSelRF} function in the \texttt{varSelRF} package such that the forests grown during the process would be conditional inference forests rather than random forests and the initial importances would come from the results of the 30 trials rather than being generated within the algorithm (and thus would allow the results to be replicated) \cite{diaz-uriarte_gene_2006, diaz2007genesrf}. We used the default value of 20\% of the variables being dropped after each trial. We use the term ``meaningful" instead of ``significant" to signify that selected factors are the factors found to provide the most information to the model and not found from a test of statistical significance. 

\section{Results\label{sec:results}}
\subsection{Model Validation}
\begin{figure}
  \includegraphics[width=\linewidth]{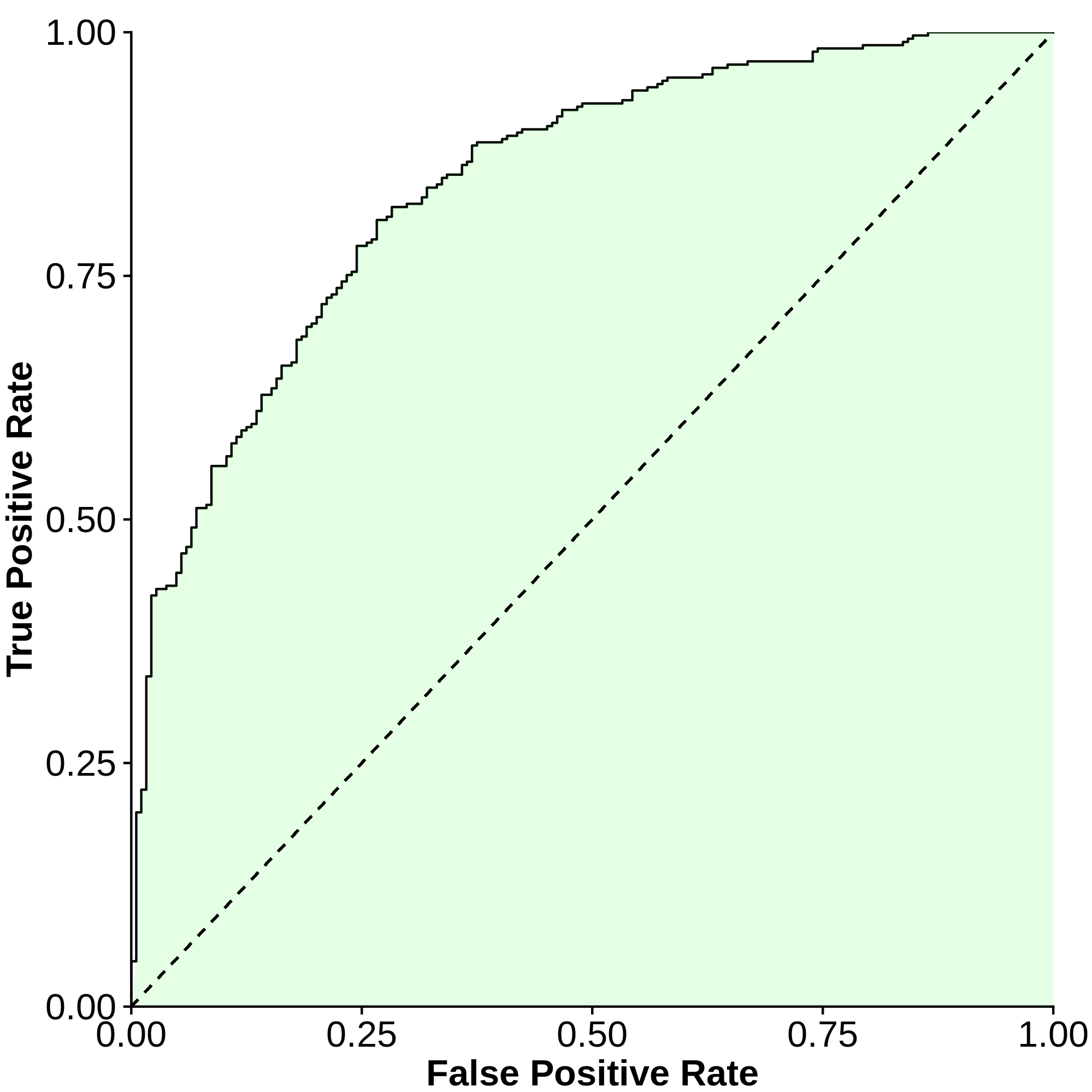}
  \caption{Representative ROC curve for one of the 30 trials of predicting whether faculty do or do not have experience teaching computation from 44 predictor variables. The AUC for this trial 0.8250, suggesting a good model as the AUC is greater than 0.7.}
  \label{fig:roccurve}
\end{figure}

Across the 30 trials, our model successfully predicted whether a faculty member had experience teaching computation 77.4\% $\pm$ 0.5\% of the time and had an AUC of 0.838 $\pm$ 0.002 (see representative ROC curve in Fig.~\ref{fig:roccurve}).  As 62.2\% of the sample had experience with teaching computation, our accuracy is significantly (both in the practical and statistical sense) higher than the non-information rate, which is the accuracy if the model were to predict every data point as belonging to majority class, in our case, the faculty with experience teaching computation. From the confusion matrix shown in Table \ref{tab:confusionmatrix}, we see that the model is better at predicting faculty with experience teaching computation compared to faculty without experience teaching computation. This difference in prediction ability may be caused by the fact that there are approximately 50\% more faculty with experience teaching computation than faculty without experience teaching computation; we address this further in Sec.\ref{sec:limitations}. Since our accuracy is significantly higher than the non-information rate and the AUC is above 0.8, our model can satisfactorily predict whether a faculty member has experience teaching computation.

% Just be lazy and use https://www.tablesgenerator.com/ to make the header with multiple lines -Nick
\begin{table}
\centering
\begin{tabular}{lllll}
\multicolumn{2}{l|}{\multirow{2}{*}{\begin{tabular}[c]{@{}l@{}}Do you have experience \\ teaching computation?\end{tabular}}} & \multicolumn{2}{l}{Data Says} &  \\
\multicolumn{2}{l|}{} & Yes & No &  \\ \cline{1-4}
\multirow{2}{*}{\begin{tabular}[c]{@{}l@{}}Model\\  Predicts\end{tabular}} & \multicolumn{1}{l|}{Yes} & \textbf{56.7\%} & 16.8\% &  \\
 & \multicolumn{1}{l|}{No} & 5.8\% & \textbf{20.7\%} &  \\
 &  &  &  & 
\end{tabular}
   \caption{Confusion matrix for a representative trial of the 30 trials. Numbers in bold are correct predictions and add to the accuracy} \label{tab:confusionmatrix}
\end{table}

\subsection{Feature Importance}
\subsubsection{Features that are more important}

\begin{figure*}
  \includegraphics[width=\linewidth]{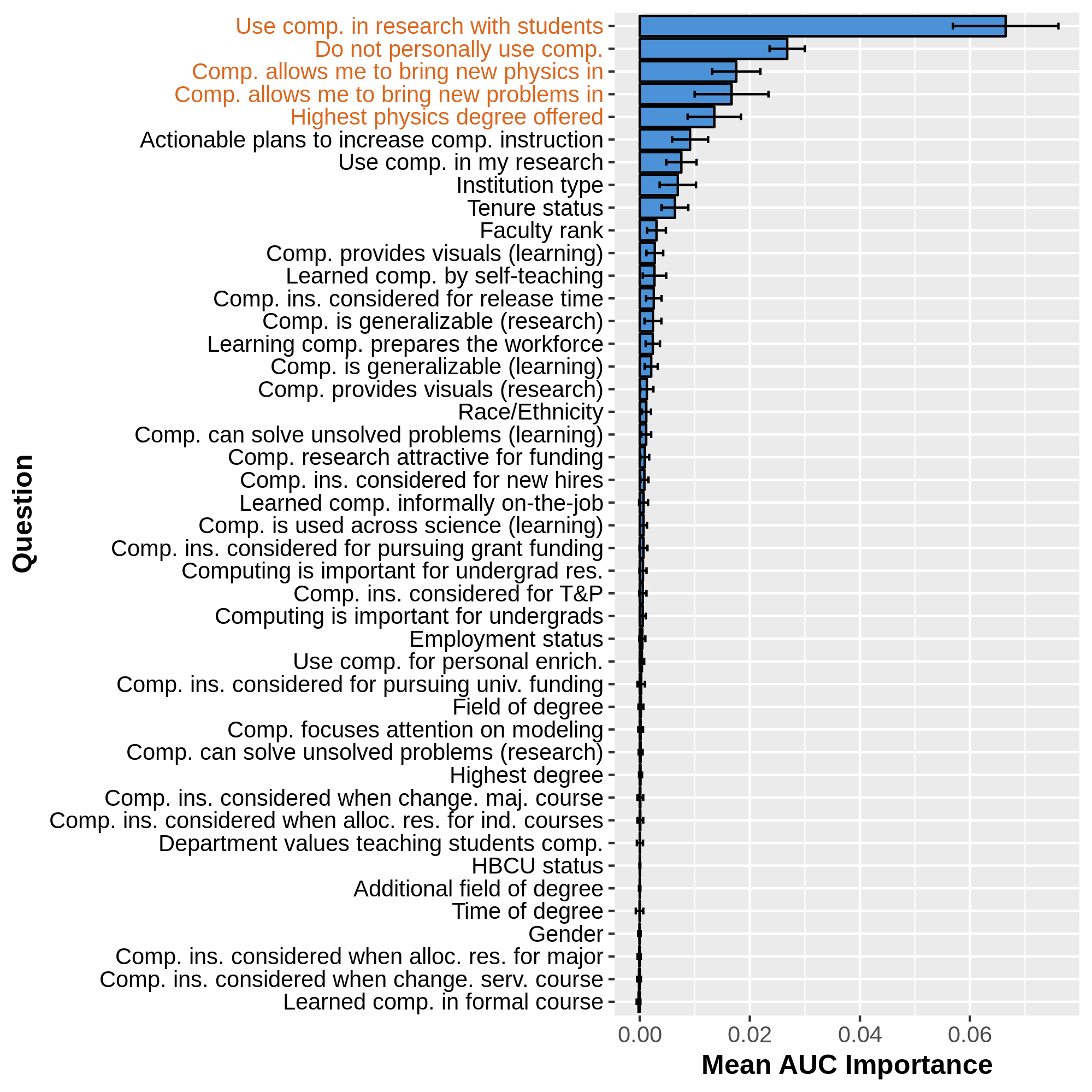}
  \caption{Variable importances for each of the 44 factors used to predict whether a faculty member has experience teaching computation. The importance is based on the average change in the AUC if the factor is permuted. Thus, factors that change the AUC the most have the largest importances. The first five factors in color are the ones selected from the recursive backward elimination approach. The error bars represent a standard error of the mean AUC importance. Full questions can be found in the appendix.}
  \label{fig:impplot}
\end{figure*}

In addition to generating predictions, our model is able to determine the importance of each variable that was used in predicting whether a faculty has experience teaching computation; these importances are shown in Fig.~\ref{fig:impplot}. The variable importances here are computed using AUC-based permutations methods, meaning the importance shown in the plot is the average decrease in the AUC if the variable were permuted and its association with the response variable were broken. For example, if the responses in the variable ``Use comp. in research with students'' were randomly shuffled, the AUC in Fig. \ref{fig:roccurve} would drop from 0.825 to approximately 0.760, a 0.065 decrease which is that variable’s importance as shown in Fig.~\ref{fig:impplot}.

We find that the most important features are ``I use computation in research with students'', ``I do not personally use computation'', ``computation allows me to bring new physics into my classroom'', ``computation allows me to bring new problems into my classroom'', and the highest physics degree offered by the institution. Actionable plans to increase computational instruction,  ``I use computation in my research'', institution type, and tenure status are slightly less important features for predicting whether faculty have experience teaching computation. When we perform the recursive backward elimination technique, we find that the meaningful features are ``I use computation in research with students'', ``I do not personally use computation'', ``computation allows me to bring new physics into my classroom'', ``computation allows me to bring new problems into my classroom'', and the highest physics degree offered by the institution.

\begin{figure*}
  \includegraphics[width=\linewidth]{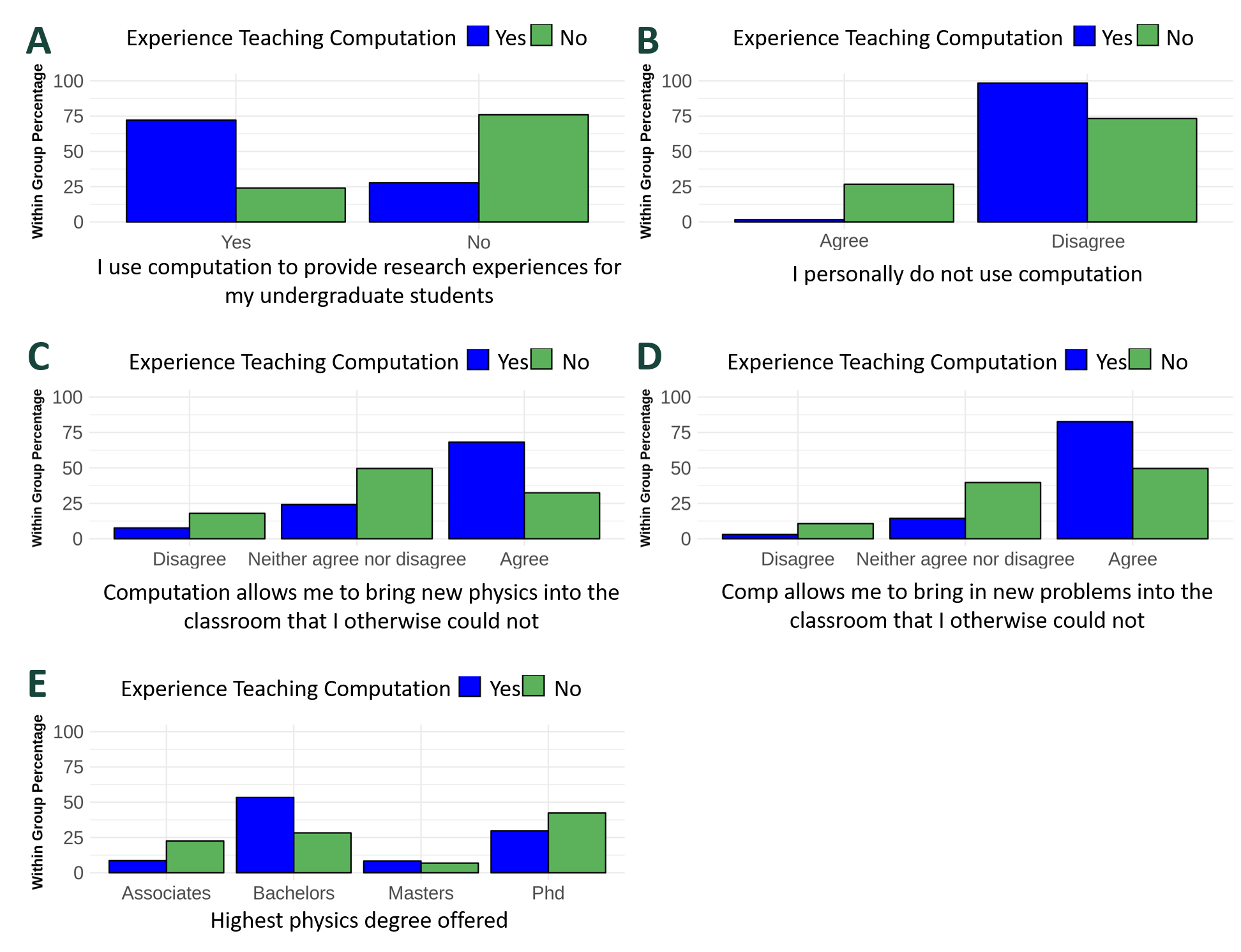}
  \caption{Distribution of responses based on whether faculty do or do not have experience teaching computation. Here, within group percentage means the percentage of faculty within the group who use computation or the group who does not use computation. Since these five factors are the meaningful factors, we expect that the distribution of responses should be different between faculty with experience teaching computation and those who do not. Plot A shows the distribution of the most important feature while plot E shows the fifth most important feature. All five plots have $\chi^2$ with corrected $ \textit{p} < 0.05$.}
  \label{fig:posplots}
\end{figure*}

As a check that these five meaningful variables are in fact meaningful, we then reran the $\texttt{cforest}$ algorithm 30 times with just these five variables. We obtained an average accuracy of 76.4\% $\pm$ 0.4\% and an AUC of 0.818 $\pm$ 0.002. Recall that the values obtained when using all 44 variables were 77.4\% $\pm$ 0.5\% and 0.838 $\pm$ 0.002 respectively.
As the accuracies are nearly the same and the AUC of the five meaningful variables model is still above 0.8, we can further support the claim that these five variables are meaningful.

While the importances are useful for determining which variables are good discriminators between faculty who have experience teaching computation and faculty who do not, the importances by themselves cannot say which group is more likely to have a specific trait. To determine which group is more likely to possess a specific trait, the distributions of responses must be investigated. The distributions of the five meaningful variables are shown in Fig.~\ref{fig:posplots}. For example, faculty who have experience teaching computation tend to use computation to provide undergraduate students with research experience while faculty who do not have experience teaching computation tend not to personally use computation. Similarly, faculty who have experience teaching computation tend to agree that computation allows them to bring new physics and new problems into the classroom that would not be possible without using computation to a greater degree than those who do not have experience teaching computation.

\subsubsection{Features that are less important}

\begin{figure*}
  \includegraphics[width=\linewidth]{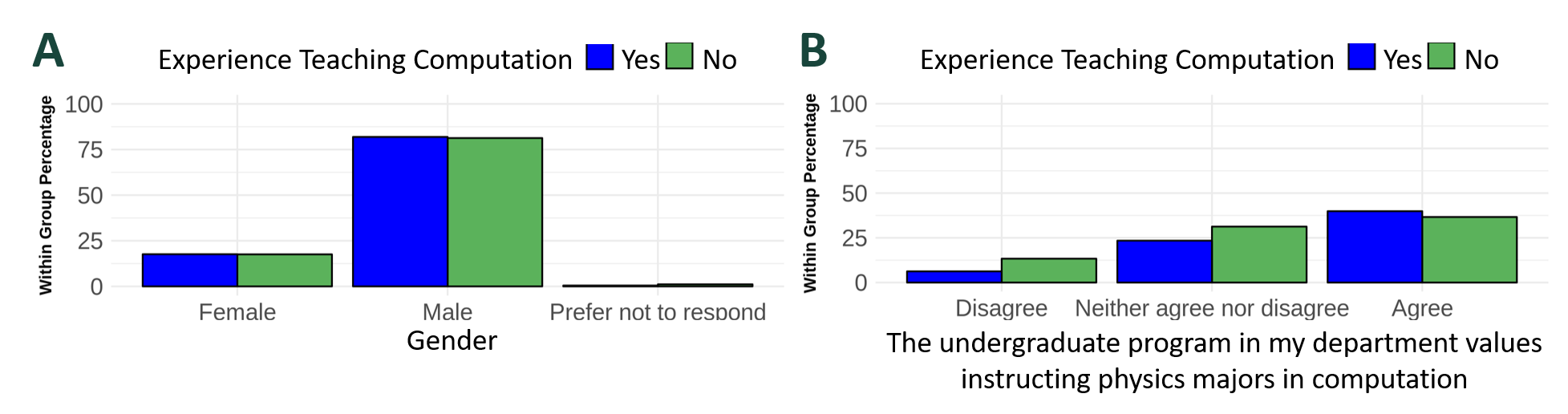}
  \caption{Distribution of responses based on whether faculty do or do not have experience teaching computation. Here, within group percentage means the percentage of faculty within the group who use computation or the group who does not use computation. Since these two variables are less useful for predictions, we expect that the distribution of responses should be not be different between faculty with experience teaching computation and those who do not.}
  \label{fig:negplots}
\end{figure*}

In addition to looking at which features are good discriminators between faculty who do and do not have experience teaching computation, investigating which features are not as good discriminators can be informative. For example, we find that demographic factors such as race/ethnicity, gender, time of degree (a proxy for age), field of degree, and highest degree obtained are among the less important factors. In addition, we find that departmental and institutional factors (statements in Fig.~\ref{fig:impplot} that begin with ``comp. ins. considered...'') are also among the less important factors. The distributions of some of these factors are shown in Fig. \ref{fig:negplots}. Compared to the distributions of features that are more important, the features that are less important seem to appear equally among the faculty who do and do not have experience teaching computation. These differences in distributions provide further support that the five meaningful factors are indeed meaningful.

\section{Validating our choice of hyperparameters}
Random forest and conditional inference forest models have multiple parameters that can be adjusted to control how the forest grows. As these parameters need to be picked before the model is created, they are called hyperparameters and choices for these hyperparameters can affect the quality of the forest grown. For example, if the amount of the data from which the forest is grown (training fraction) is increased, the predictions should improve up to some threshold. Likewise, if the number of trees in the forest is increased, the quality of the predictions should increase up to some threshold. In this section, we assess the stability of our model by varying the training fraction and the number of trees in the each forest. If our findings do not vary significantly as the training fraction and number of trees vary, we can be more confident that our results actually are representative of the data and are not artifacts of the model. As we are more concerned with identifying the important factors than the predictive power of the model, we do not perform a grid search to identify the set of hyperparameters that would result in the highest accuracy or area under the curve. Prior work has found that randomly choosing hyperparameters is more efficient and provides comparable results to a typical grid search \cite{bergstra_random_2012}. 

\subsection{Effects on accuracy and area under the curve}

\begin{figure}
  \includegraphics[width=\linewidth]{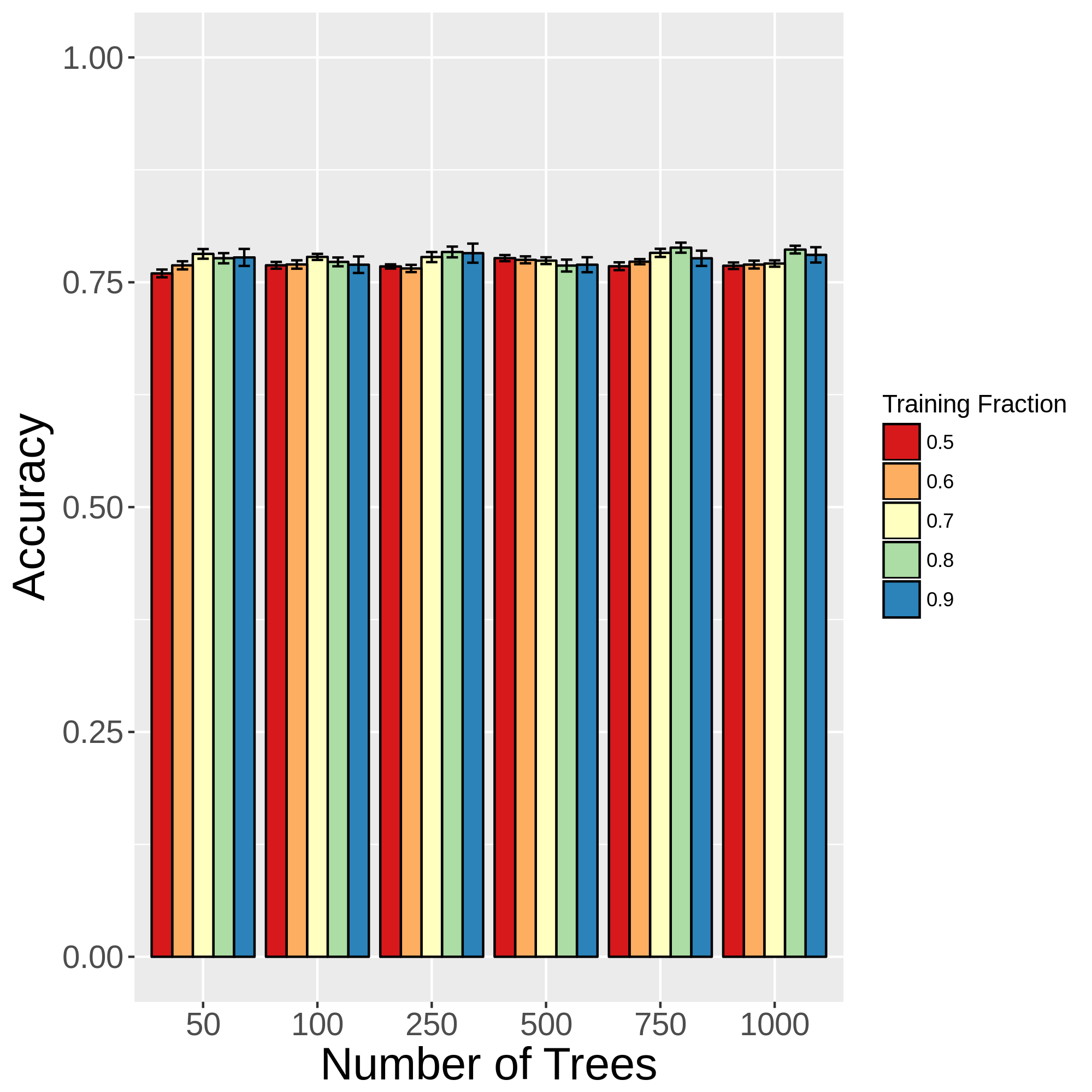}
  \caption{Average accuracy of the model for various training fractions and number of trees in the forest. Error bars correspond to a standard error.}
  \label{fig:accplots}
\end{figure}

\begin{figure}
  \includegraphics[width=\linewidth]{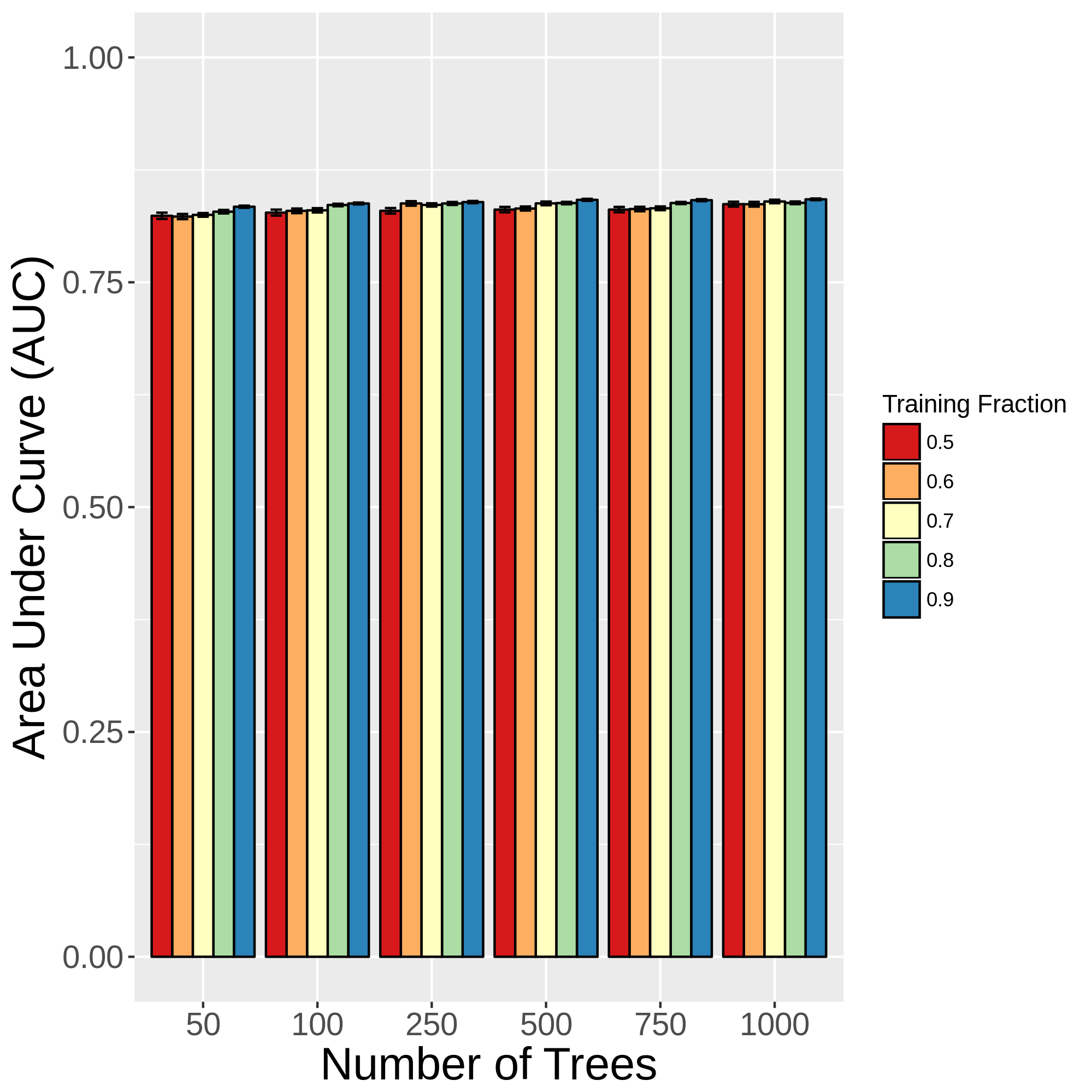}
  \caption{Average area under the curve of the model for various training fractions and number of trees in the forest. Error bars correspond to a standard error.}
  \label{fig:aucplots}
\end{figure}

To check for variation, we selected five training fractions, 0.5 (split the data in half), 0.6 (used when creating a training, validation, and testing set), 0.7 (our original choice), 0.8 (amount used for a 5-fold validation), and 0.9 (amount used for a 10-fold validation) and six forest sizes (50, 100, 250, 500, 750, 1000), where 500 trees was our original choice, informed by practical considerations and Svetnik $\textit{et al}$.'s finding that error rates stabilize on the order of $10^2$\cite{svetnik2003random}. For each pair of training fraction and forest size, we ran our cforest algorithm 30 times, creating 870 new forests (29 new models with 30 trials each). We then averaged our results across forests with the same training fraction and number of trees. The accuracy and AUC for each pair of training fraction and number of trees are shown in Fig.~\ref{fig:accplots} and Fig.~\ref{fig:aucplots} respectively. A visual inspection suggests that neither the accuracy nor area under the curve vary significantly when the training fraction and the number of trees in the forest are changed. Indeed, the ranges of the accuracy and AUC are 0.03 and 0.02 respectively, which are insignificant from a practical perspective. Thus, while this range represents multiple standard deviations, it is of little practical significance so we can be confident that our model would not significantly improve or become worse by selecting a different set of hyperparameters.

\begin{figure}
  \includegraphics[width=\linewidth]{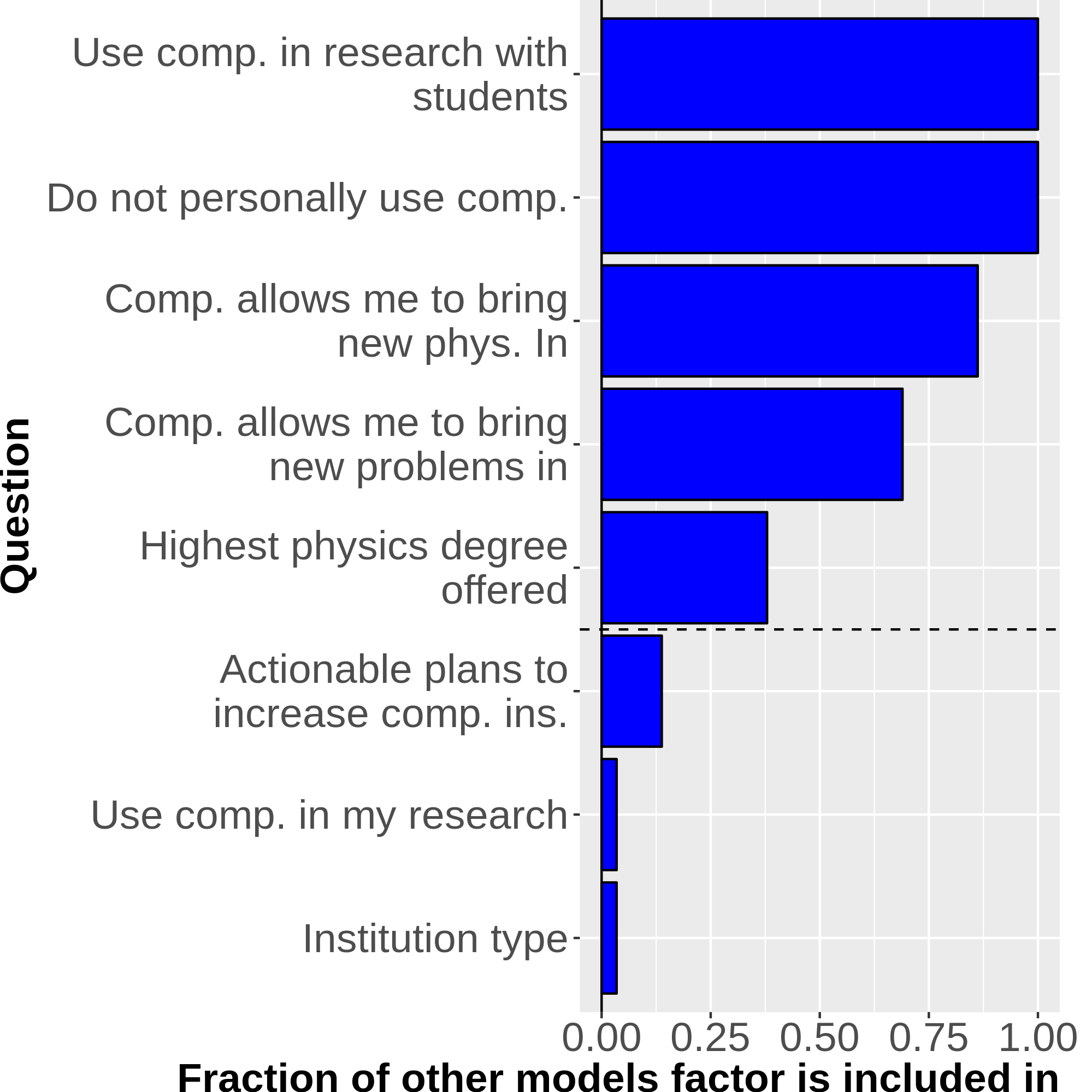}
  \caption{Fraction of the other 29 models in which the variables are selected as meaningful. Variables above the line were selected as meaningful in the original model.}
  \label{fig:ps_occplot}
\end{figure}

\subsection{Effects on variable importance}
Because there were small variations in the area under the curve for varying training fractions and number of trees, we may expect there to be variation in the variable importances as the variable importances are based on changes in the area under the curve. We expect there to be natural variation in the importances just from using different training sets, so we chose to only focus on the selected meaningful variables. We used the same choices of hyperparameter combinations as in the previous section. 
Fig.~\ref{fig:ps_occplot} shows the fraction of the 29 new models where each variable was found to be meaningful. Variables that are not shown were not found to meaningful in any of the models. We see that ``using computation in research with students'' and ``I do not personally use computation'' were selected as meaningful in all the models while ``computation allows me to bring new physics into my course'' and ``computation allows me to bring new problems into my course'' were selected as meaningful in over two-thirds of the models. Highest physics degree offered was selected as meaningful in only about a third of the other models, suggesting that the selection of this factor may be influenced by how the model is constructed or should be more accurately described as marginally meaningful. None of the other factors appeared in more than 15\% of the models. These results suggest that at least four of the selected meaningful factors are in fact meaningful and not just artifacts of how we constructed our model while highest physics degree offered may be marginally meaningful and influenced by the hyperparameters chosen.

\section{Limitations}\label{sec:limitations}
In this section, we comment on how our model may be limited based on the nature of the data and the amount of missing data.

\subsubsection{Unbalanced classes}
Because 60\% of our data is from faculty who have experience teaching computation, training data set used to grow the Random Forest model will contain more instances of faculty who have experience teaching computation than faculty who do not have experience teaching computation. As there are more instances of faculty with experience teaching computation to learn from, we expect that the model will be better at correctly classifying faculty with experience teaching computation than faculty without experience teaching computation, which we observed in our data. However, our model appears to be classifying faculty without experience teaching computation almost at random, suggesting that our results are biased. While many methods have been proposed to correct the imbalance, these methods can introduce additional biases of their own. For example, the data can be artificially balanced by bootstrapping the minority class (up-sampling), until the classes are equal. However, the cforest algorithm is not unbiased when bootstrapping is used \cite{strobl_bias_2007}. Alternatively, the data could be down-sampled, where only a random sample of the majority class equal in size to the minority class is used to grow the forest. While this does not introduce bias in the cforest algorithm, it would require excluding nearly 20\% of the usable data from the model and in our trials, did not improve the accuracy or AUC of the models. Finally, there have been alternative variations of the random forest algorithm, including balanced random forests, which is based on up-sampling, and weighted random forests, which are based on cost sensitive learning, to handle unbalanced classes. However, these algorithms are based on CART algorithms meaning the original biases conditional inference forests were designed to combat would be reintroduced. Thus, adequately handling imbalanced data without introducing or reintroducing bias appears to be an open problem.

Alternatively, if we are using our model as a truly predictive model, we would not know the true values until after the prediction. In the case that the model does predict that a faculty member does not have experience teaching computation, it would be correct about 80\% of the time ($20.7\%/(20.7\%+5.8\%)$). Thus, whether the algorithm should be viewed as biased depends on what we view as the given information, the model's prediction or the actual response.

While our predictions may contain some bias, the predictive model was not the focus of this study. Instead, we are concerned with determining which factors are discriminators between faculty with or without experience teaching computation. The factors were determined from values of the AUC importance, which is not affected by class imbalance because the area under the curve weighs the majority and minority class equally \cite{janitza_auc-based_2013}. The AUC importance is not involved in the growing of the forest however, meaning that the predictions can still be biased from unbalanced classes. Thus, we believe that our results are minimally impacted by having more faculty with experience teaching computation in our sample.

\subsubsection{Missing data}
For this paper, we decided to employ complete case analysis, which means we only included faculty who responded to all 44 questions used in the study. This left us with just over half of the original participants, meaning we excluded nearly half of the participants, which could affect our results. As described by Hapfelmeier et al, there do exist methods for random forests with missing data and generating variable importance measures \cite{hapfelmeier_new_2014}. When we implemented this approach, our accuracy decreased to $72.2\% \pm 0.4\%$ but the AUC increased to $0.889 \pm 0.001$. In terms of the selected meaningful variables, we found that the five variables selected in our initial model and one additional variable (``I used computation in my research'') to be meaningful. However, this set of meaningful variables could also be observed by varying the training fraction and the number of trees in our original model. Therefore, it does not appear that excluding faculty with missing data changes our results and hence we decided to use the complete case analysis because complete case analysis resulted in a better predictive model. 

\section{Discussion\label{sec:discussion}}
To interpret our results, we can compare the meaningful factors to those found in the literature. Of the twenty factors Henderson, Dancy, and Niewiadomska investigated using logistic regression \cite{henderson_use_2012}, four appear in our study directly: highest degree obtained, gender, type of institution, and type of position, which we called employment status. Additionally, we can relate three of their variables to three of our computation specific variables: we treat their ``department encouragement'' variable as our  ``department values teaching students computation'' factor, their ``interest in using more RBIS'' variable as our ``actionable plans to increase computation'' factor and their ``years of teaching experience'' variable as our ``time of degree'' factor as both of these can be viewed as proxies for age. As our survey was designed to cover five broad areas and to limit survey fatigue, not all of the factors from \cite{henderson_use_2012} could be included in the survey. Of these seven factors, we then expect that the factors Henderson et al. found to be correlated with trying or not trying RBIS (institution type and interest in using more RBIS) to be among the factors we found to be more predictive of a faculty member having experience teaching computation while their other five variables should be among the factors we found less important. Indeed, we found that type of institution and actionable plans to increase computation were among the more important of our factors while the other five were among the less important factors. We note however that type of institution and actionable plans to increase computation were only meaningful factors in less than 15\% of the models we created when varying the number of trees and the training fraction and were not meaningful in our original model.

As teaching and research expectations vary based on the institution type, we may expect some types of institutions allow their faculty more time to focus on their courses. For example, faculty at institutions that only offered bachelor’s degrees in physics were more likely than faculty at any other type of institution to have experience teaching computation. This may be due to lower research demands and hence, more time to devote to developing and preparing their courses. Therefore, faculty at these types of institution may have already overcome one of the implementation challenges, having time to do so. Likewise, those who have already made plans to integrate computation into their courses have overcome the challenge of fitting more material into their courses.

On the other hand, factors such as a faculty member’s gender, highest degree obtained, and years of teaching experience do not address any of the implementation challenges so we do not expect these to be important factors. Departmental encouragement and type of position may indirectly relate to an implemental challenge such as approving changing curriculum to accommodate computation or by creating time to work on implementing computation. However, type of position only refers to full-time, part-time, or course-by-course, not the actual duties of the position, so it is unlikely that this factor provides much information about time for implementing computation beyond the number of hours worked each week. As these two factors are at most indirectly related to challenges with implementing computation or a RBIS, it seems reasonable that they are not important factors for discriminating between faculty who do and do not have experience during computation.

%talk about early adopters from Rogers: http://sphweb.bumc.bu.edu/otlt/MPH-Modules/SB/BehavioralChangeTheories/BehavioralChangeTheories4.html

%https://teddykw2.files.wordpress.com/2012/07/everett-m-rogers-diffusion-of-innovations.pdf
One reason that we may not be finding the same important factors as found in the literature for adopting RBIS is that the faculty who are using computation are likely what Rogers calls the early adopters \cite{rogers1995diffusion}. The literature on adopting RBIS focuses around more established instructional strategies and hence the early and late majorities. We would expect that the early adopters of a new instructional strategy would be those familiar with the strategy and see a clear benefit to using the strategy instead of continuing to the use the strategies they had previously been using. This is the pattern we observe in our meaningful factors: those who use computation in their research with students tend to use computation while those who do not personally use computation tend not to have experience teaching computation. Likewise, those who believe computation allows new physics and new problems to be incorporated into the curriculum, a clear benefit of using computation, are more likely to have experience teaching computation.

As 60\% of the respondents to our survey indicated they have experience teaching computation, our claim that these faculty are early adopters may seem contentious as 60\% is a majority of faculty. However, Caballero and Merner note that those who use computation are more likely to respond to the survey than those who do not use computation \cite{caballero_prevalence_2018}. Thus, 60\% should be thought of as an upper limit on the percentage of faculty using computation in their courses. 

Regardless of how we classify these faculty, it is important to note that these results are just a snapshot of the state of computation now. As computation in the classroom becomes adopted by more physics faculty, we expect that these meaningful factors will change and will likely more closely align with factors correlated with trying a RBIS. Currently though, the important factors were focused on the what the individual does, using computation in research with students or not using computation personally, or believes, computation adds new physics and problems to the course, and not on institutional or departmental factors, suggesting that integrating computation into a course is a personal choice, which does align with previous findings that faculty adopt new instructional strategies based on their own decisions.

\section{Conclusion and Implications\label{sec:conclusions}}
In this paper, we created a random forest model to predict whether physics faculty have experience teaching computation. From our model, we find four meaningful factors and one marginally meaningful factor that discriminate between faculty who do and do not have experience teaching computation: using computation in their research with students, not personally using computation, believing computation allows them to bring new physics into their course, believing computation allows them to bring new problems into their course, and the highest physics degree offered at their institution. Since most of the meaningful factors are related to faculty choice and there is lack of institutional or department factors, we conclude that deciding to teach computation is viewed as a choice by physics faculty members.

As the meaningful factors were at the individual level instead of the departmental or institutional level, the implications of our study are then that at this moment, efforts to increase computation use should be at the level of individuals rather than at a departmental level. If we do characterize those who use computation as early adopters, then future work should focus on the faculty who will make up the early majority, which need to see evidence that computation adds value to their course before they will adopt it \cite{rogers1995diffusion}. Broadly, the meaningful factors suggest that faculty who have experience using computation and see value in teaching computation will do so while those who do not use computation in their professional work or do not see how computation can complement their course's current content will not teach computation. These findings are perhaps not surprising as a study outlining a vision for integrating computation into undergraduate physics courses concludes that integrating computation will require ``many faculty minds to change" and many ``faculty skills to train" \cite{Chonacky:2008gq}. The fact that these factors are still relevant a decade later suggest there is still a long way to go to widespread implementation of computation in undergraduate physics courses. 

\begin{acknowledgements}
The authors would like to thank Laura Merner, Norman Chonacky, and Robert Hilborn for their work to develop survey areas. The authors would also like to thank members of PERL@MSU for their helpful comments on drafts of this paper. This work was supported by the National Science Foundation's Division of Undergraduate Education (DUE-1431776 and DUE-1432363) and by Michigan State University.
\end{acknowledgements}

\section{Appendix}\label{sec:appendix}
In the appendix, we provide the wording of the questions we used in the survey and the shortened versions of those questions that are used throughout the paper.

%\newpage

\bibliography{refs,Nick_ref,HD_other_refs}
\bibliographystyle{apsper}

\setcounter{table}{0}
\newpage
\newpage
\onecolumngrid
\setlength\LTleft{-3cm}
\setlength{\tabcolsep}{5pt}

\begin{longtable}{p{6cm}p{10cm}}
\caption{Full list of survey questions}\\
\hline \multicolumn{1}{c}{\textbf{Shortname}} & \multicolumn{1}{c}{\textbf{Question Statement}} \\ \hline
\endfirsthead
Teaching Computation & Given this broad definition of computation, do you have any experience teaching computation to undergraduate physics students? \\
Computing is important for undergrad res. &  Rate the degree to which you agree or disagree with the following statements. For the following questions we would like to understand your personal perspective of the role of computation in physics. - I think that computation is important for undergraduate physics research. \\
Computing is important for undergrads &  Rate the degree to which you agree or disagree with the following statements. For the following questions we would like to understand your personal perspective of the role of computation in physics. - I think that learning computation is important for undergraduate physics majors. \\
Department values teaching students comp. &  Rate the degree to which you agree or disagree with the following statements. For the following questions we would like to understand your personal perspective of the role of computation in physics. - The undergraduate program in my department values instructing undergraduate physics majors in computation. \\
Comp. can solve unsolved problems (research) & With regard to your personal research, rate the degree to which you agree or disagree with the following statements: - Computation can solve unsolvable (analytical) problems. \\
Comp. is generalizable (research) & With regard to your personal research, rate the degree to which you agree or disagree with the following statements: - Computation is generalizable to many different kinds of problems. \\
Comp. provides visuals (research) & With regard to your personal research, rate the degree to which you agree or disagree with the following statements: - Computation affords visualization (graphs, animations) of solutions. \\
Comp. research attractive for funding & With regard to your personal research, rate the degree to which you agree or disagree with the following statements: - Computational research is attractive to funding agencies. \\
Comp. is used across science (learning) & Rate the degree to which you agree or disagree with the following aspects of learning computation: - Computation is used in many science and engineering applications. \\
Comp. can solve unsolved problems (learning) & Rate the degree to which you agree or disagree with the following aspects of learning computation: - Computation can solve unsolvable (analytical) problems. \\
Comp. is generalizable (learning) & Rate the degree to which you agree or disagree with the following aspects of learning computation: - Computation is generalizable to many different kinds of problems. \\
Comp. provides visuals (learning) & Rate the degree to which you agree or disagree with the following aspects of learning computation: - Computation affords visualization (graphs, animations) of solutions. \\
Comp. focuses attention on modeling & Rate the degree to which you agree or disagree with the following aspects of learning computation: - Computation focuses student's attention on modeling the important physics of a problem. \\
Learning comp. prepares the workforce & Rate the degree to which you agree or disagree with the following aspects of learning computation: - Learning computation prepares students for the modern scientific workforce. \\
Comp. allows me to bring new physics in & Rate the degree to which you agree or disagree with the following aspects of learning computation: - Computation allows me to bring new physics into the classroom that I otherwise couldn't. \\
Comp. allows me to bring new problems in & Rate the degree to which you agree or disagree with the following aspects of learning computation: - Computation allows me to bring new problems into the classroom that I otherwise couldn't. \\
Comp. ins. considered for new hires & What level of consideration does your department give to undergraduate instruction in computation when making decisions regarding: - Hiring new faculty members \\
Comp. ins. considered for T\&P & What level of consideration does your department give to undergraduate instruction in computation when making decisions regarding: - Tenure and promotion decisions \\
Comp. ins. considered when alloc. res. for major & What level of consideration does your department give to undergraduate instruction in computation when making decisions regarding: - Allocating resources for programs or tracks within the undergraduate major \\
Comp. ins. considered when alloc. res. for ind. courses & What level of consideration does your department give to undergraduate instruction in computation when making decisions regarding: - Allocating resources for individual undergraduate courses \\
Comp. ins. considered when change. serv. course & What level of consideration does your department give to undergraduate instruction in computation when making decisions regarding: - Changing undergraduate service courses \\
Comp. ins. considered when change. maj. course & What level of consideration does your department give to undergraduate instruction in computation when making decisions regarding: - Changing courses for undergraduate majors \\
Comp. ins. considered for release time & What level of consideration does your department give to undergraduate instruction in computation when making decisions regarding: - Releasing time for faculty to develop computation in undergraduate courses \\
Comp. ins. considered for pursuing univ. funding & What level of consideration does your department give to undergraduate instruction in computation when making decisions regarding: - Pursuing university funding \\
Comp. ins. considered for pursuing grant funding & What level of consideration does your department give to undergraduate instruction in computation when making decisions regarding: - Pursuing grant funding \\
Actionable plans to increase comp. instruction & Do you have concrete and actionable plans to increase your use of computation in your own undergraduate physics teaching in the next year? \\
Highest degree & What is your highest degree? \\
Field of degree & In what field did you receive your highest degree? \\
Additional field of degree & In what other field did you receive your highest degree? \\
Time of degree & When did you obtain your highest degree? \\
Learned comp. by self-teaching & How did you come to learn computation? - Self-taught \\
Learned comp. informally on-the-job & How did you come to learn computation? - Informal on-the-job \\
Learned comp. in formal course & How did you come to learn computation? - Formal course(s) \\
Use comp. for personal enrich. & How do you personally use computation? - Exclusively for personal enrichment/use  \\
Use comp. in my research & How do you personally use computation? - In my research work \\
Use comp. in research with students & How do you personally use computation? - To provide research experiences for my undergraduate students  \\
Do not personally use comp. & How do you personally use computation? - I do not use computation.  \\
Faculty rank & What is your current faculty rank? \\
Employment status & As of the Spring 2016 term, what was your employment status \\
Tenure status & Are you currently a tenured faculty? \\
Gender & What is your gender? \\
Race/Ethnicity & What is your race or ethnicity? \\
HBCU status & Is your institution a historically black college or university? \\
Institution type & What type of institution do you work at? \\
Highest physics degree offered & What is the highest physics degree offered at your institution? \\
\hline
\hline
\end{longtable}

\end{document}